\documentclass[twocolumn,a4paper,10pt]{svjour3}          % twocolumn
\smartqed

\usepackage{graphicx}
\graphicspath{ {./Visualisations/} }

\usepackage{microtype}

\AtBeginDocument{%
  \providecommand\BibTeX{{%
    \normalfont B\kern-0.5em{\scshape i\kern-0.25em b}\kern-0.8em\TeX}}}

\usepackage{xspace}
\usepackage{paralist}
\newcommand{\eg}{e.\,g.,\xspace}
\newcommand{\ie}{i.~e.,\xspace}
\newcommand{\cf}{cf.\xspace}

\usepackage{url}

\begin{document}

\title{Social Analytics of Team Interaction using Dynamic Complexity Heat Maps and Network Visualizations}

\author{Travis J. Wiltshire, Dan Hudson, Philia Lijdsman, Stijn Wever  and \mbox{Martin Atzmueller}
}

\institute{
T. J. Wiltshire and D. Hudson \at Tilburg University, Department of Cognitive Science and\\ Artificial Intelligence, Tilburg, The Netherlands\\
\email{\{T.J.Wiltshire,d.d.hudson\}@uvt.nl}
\and P. Lijdsman and S. Wever \at Bricklayers V.O.F., Baarn, The Netherlands\\
\email{\{philia,stijn\}@bricklayers.nl}
\and M. Atzmueller \at Osnabr\"uck University, Institute of Computer Science,\\ Semantic Information Systems Group, Osnabr\"uck, Germany\\
\email{martin.atzmueller@uni-osnabrueck.de}
}

\date{}

\maketitle

\begin{abstract}
Given the increasing complexity of many sociotechnical work domains, effective teamwork has become increasingly crucial. While there is evidence that face-to-face communication contributes to effective teamwork, methods for understanding the time-varying nature and structure of team communication are limited.
In this work, we combine sensor-based social analytics of Sociometric badges (Rhythm Badge) with two visualization techniques (Dynamic Complexity Heat Maps and Network Visualizations) to advance an intuitive way of understanding the dynamics of team interaction.
To demonstrate the utility of our approach, we provide a case study that examines one team's interaction for a Lost at Sea simulation. We were able to recover transitions in the task and team interaction as well as uncover structural changes in team member energy and engagement, which we visualize using networks. Taken together, this work represents an important first step at optimizing team effectiveness by identifying critical transitions/events in team communication and interaction patterns.
\end{abstract}


\section{Introduction}
Advances in technological innovation often give rise to an increase in the complexity of work in many sociotechnical domains such as business, aerospace, healthcare, and science. Rises in complexity of work often bring about increased cognitive demands on individuals and, thus, teamwork is increasingly crucial. While communication contributes to effective teamwork \cite{marlow2018does}, understanding of the multi-modal and multi-scale nature of team communication is limited \cite{gorman2016cross,dale2013self}. That is, mutual information exchange facilitates team coordination across many modalities including verbal and non-verbal, and this coordination can vary over time and behave differently at different time scales \cite{oullier2009social,wiltshire2019multiscale}. 

To gain a better understanding of the role of team communication as it changes over time, in this work, we employ a combined sensor-based social analytics approach~\cite{kibanov2019social} utilizing the Rhythm Badge platform~\cite{lederman2018rhythm}, which has been shown to be effective for capturing social interactions~\cite{Atzmueller:14:CoRR} with promising, yet under-researched, applications to team science~\cite{kozlowski2018unpacking}. Therefore, we also employ two visualization techniques of the sensor data: Dynamic Complexity Heat Maps \cite{schiepek2017monitoring} and Network Visualizations \cite{paxton2013multimodal}. The former quantifies critical moments of change in the team interaction and the latter is used to show the strength of the communicative connections between team members during an interaction. We apply our approach to a case study in order to demonstrate its potential as an analytic tool. To the best of our knowledge, this combination of sensors and visualizations has not been researched previously.
%Thus, in the section that follows, we discuss our theoretical framework and related work. 

\vspace*{-.3cm}

\section{Related Work}

Below, we first discuss related theory and work on team dynamics, Sociometric badges, and network visualizations. After that, we detail the relevance of critical instabilities for studying changes in team work.

\subsection{Team Dynamics}
Teams typically consist of two or more individuals that work independently toward a common goal \cite{salas1992toward}. As teams work on increasingly complex tasks and novel problems, understanding the dynamics of collaborative team processes and performance is essential \cite{fiore2010towards,fiore2010toward}. While team research has spanned several decades \cite{salas1992toward}, only a small percentage of that work has focused on dynamics of teams as complex adaptive systems \cite{ramos2018teams}. Thus, from a theoretical perspective, our work is motivated by the dynamical systems theory approach to teams \cite{gorman2017understanding}. 

A dynamical systems approach recognizes teams are comprised by many interacting components (e.g., team members with each other and their technology) that are embedded within other systems (e.g., organizations) and where such interactions give rise to complex patterns that are changing over time and in ways that can serve adaptive functions, \cf~\cite{gorman2017understanding,ramos2018teams,cooke2013interactive}. By considering teams as dynamical systems, we can understand the temporal evolution of team and task work behaviors as well as  cognitive processes as they are evidenced in the interaction amongst team members and their technology \cite{cooke2013interactive,fiore2016technology}. While meaningful team dynamics have been observed in physiology \cite{palumbo2017interpersonal,kazi2019team}, movements \cite{wiltshire2019multiscale}, and other interpersonal aspects such as team formation and composition that change over time \cite{delice2019advancing}, in this case, we focus specifically on the interactive aspects that are observable through sensing technologies that are related to team communication. 

Much of the work examining team communication has focused on summary statistics such as the quality and frequency of certain communication types \cite{marlow2018does}. However, other work examining team communication dynamics has focused on how the coordination patterns vary over time and are related to task performance under varying conditions \cite{gorman2019measuring}. From a systems approach, we define \textit{coordination} as the ways in which system components and processes change together over time \cite{butner2014modeling}. With regard to coordination of communication, Gorman and colleagues \cite{gorman2010team}, for example, developed a method for examining how particular types of team communication processes (i.e., information, negotiation, and feedback) are coordinated if they occur in a specific temporal order (independent of which team member said it) and found that the stability of this coordination pattern was related to overcoming more task perturbations. This notion of the stability of team communications is crucial as it can reflect the degree to which teams are performing too rigidly (i.e., if too stable) or whether they are interacting in the more adaptive and flexible manner required for complex work \cite{gorman2012measuring}. 

Coordination in communication can take other forms such as the temporal regularity of not only semantic content \cite{strang2012examining}, but also acoustic vocal properties \cite{wieder2020investigating}, word usage \cite{fischer2007linguistic}, speech rate \cite{eloy2019modeling}, as well as topics and concepts \cite{angus2011conceptual}. The challenge with using semantic and/or linguistic-based methods is that they typically are more computationally and/or time intensive. While progress is being made in this area, especially for the combined use of linguistic and acoustic features \cite{murray2018predicting}, this challenge makes these measures difficult to scale to teams of larger sizes and difficult to utilize for near-real time adaptive systems that could be used to improve team work \cite{gorman2019measuring,wiltshire2020challenges}. Despite efforts to make linguistic and semantic methods more efficient, the use of Sociometric badges affords a relatively straightforward and scale-able way to examine speaker-only and vocal coordination \cite{gorman2012measuring,strang2012examining}, especially for larger teams that may not be co-located.

\subsection{Sociometric Badges}
Our research uses sociometers to record social interaction data that are then visualized for understanding the team dynamics. Wearable sensors to measure social activity, also known as  'sociometric badges' or ‘sociometers’, have now been used by researchers for over a decade to investigate various aspects of group behavior, to study such diverse phenomena as: collaborative innovation \cite{olguin2010assessing}, patient recovery times in intensive care units \cite{olguin2009capturing}, outcomes in speed dating \cite{pentland2004social}, and the social lives of primates \cite{gelardi2020measuring}. They are noted for their ability to make accurate measurements at a large scale and over long durations, while remaining affordable to use \cite{waber2008understanding}. Sociometers automatically collect quantitative data, thus giving rise to the possibility of automated analysis of social behavior, which was not readily accessible to team research in the past. 
Sociometers are wearable devices fitted with sensors that can include Bluetooth, RFID and infrared technologies for detecting proximity between individuals, microphones to detect nonverbal vocal activity, and/or accelerometers for detecting physical activity and energy levels \cite{muller2019using,waber2008understanding}. If each member of a group is equipped with such a device, then a data set extending over time, multiple individuals, and multiple modalities can be collected. This means that sociometers provide large amounts of information that researchers can use to investigate social behavior. Over time, several designs for sociometers have been developed, with an early example being presented by Choudhury and Pentland, who also coined the term ‘sociometer’ \cite{choudhury2002sociometer}. Two prominent examples \cite{muller2019using} of the latest generation of devices are the OpenBeacon proximity tag \cite{Cattuto:2010}, from the SocioPatterns Collaboration\footnote{\url{http://www.sociopatterns.org}}, and the Rhythm Badge \cite{lederman2018rhythm} from Massachusetts Institute of Technology (MIT)\footnote{\url{https://www.media.mit.edu/posts/rhythm-badge/}}, which is used in this research. 

Sociometers have been used in a number of studies to understand teamwork, where past studies have endeavoured to relate nonverbal social behavior to a variety of alternative measurements of team performance. For example, in the context of healthcare, Olguin, Gloor and Pentland \cite{olguin2009capturing} gathered data from 67 nurses working together in a post-anaesthesia care unit, using sociometric badges over a period of 27 days. Using an array of features extracted from the recordings, they were able to achieve an $R^2$ score of .73 when predicting patient recovery time, and could predict nurses' own perceptions of stress and productivity with $R^2$ scores of .77 and .63 respectively. Applying similar methods at a one-week entrepreneurship event, Olguin and Pentland \cite{olguin2010assessing} collected on average more than two hours of data for each of 109 participants. They were able to use the sociometric data to predict whether teams would produce successful business plans to an accuracy of roughly 90\%. Zhang et al. \cite{teamsense2018} focused on the use of sociometers to measure affective states and team cohesion in order to create a method for supporting long-duration spaceflight missions. They aimed to classify the participants' perceptions of team cohesion as either 'negative' or 'not negative' on each day of a simulated space mission and were able to classify task cohesion with around .75 accuracy compared to a baseline of roughly .60 and  social cohesion with an accuracy of .65 compared to a baseline of .50. 

Other work has addressed the use of multimodal data to study social behaviours involved in teamwork without specifically packaging the sensors in a sociometric badge. Neubauer et al. \cite{neubauer2016getting} presented an experiment in which two-person teams tried to defuse a simulated bomb and investigated the impact of an 'ice-breaking' session before the task. The data was recorded with microphones, video cameras and electrocardiograms. The results suggested that verbal expressions associated with ‘sociability’, ‘cognitive processes’ and ‘insight’ were more common amongst the teams who had an ice-breaking session, and they were also more likely to display facial expressions with positive affect and less likely to display facial expressions with negative affect. More recently, Avci and Aran \cite{avci2016predicting} looked at a vast array of sociometric and psychometric measures in order to predict the outcome of a group task. Their work investigated social signals as a way to predict the performance of a group that must make decisions together, using the ELEA corpus of 40 recordings of teams performing a task where they must rank the value of a list of items given a disaster survival scenario, much like the task used in our case study. Evaluating these features as predictors, the authors were able to provide some limited evidence that the following were related to high team scores: a leader with a high score during the individual phase of the task; more silence; and, more one-directional (unreciprocated) gaze. These past results indicate that sociometric methods can be effective for studying teamwork in different contexts and with different measures of performance. Informed by the successes of such methods in previous studies of team performance, we present an approach to visualizing teamwork based on sociometric data gathered by sociometers. 

One important detail to note about previous attempts to understand team performance is that the analysis of sociometric data typically leads to a set of features that describe social behavior over the span of a single or multiple days, despite the fact that social behavior can be described at finer temporal resolutions. The reason for this is mainly that past work has used performance measures that are calculated only once per day, and so daily descriptions of social behavior were most appropriate for prediction. Nevertheless, in the context of dyadic, rather than team interactions, Curhan and Pentland showed that 'thin slices' of sociometric data spanning five minutes can be predictive of outcomes in salary negotiations and in speed dating \cite{curhan2007thin}, suggesting that short periods of behavior can reliably exhibit distinctive and important social details. Looking at even smaller timescales, Kim et al. \cite{kim2012awareness} investigated the possibility of real-time feedback in team meetings, and argued that it can have a positive impact on the way team members work together. In our research, we consider features indicative of team interaction over short durations in the hope that this will provide a more precise and detailed way to investigate teamwork and performance. 

There are numerous ways that it might be possible to choose the temporal scale at which to analyze sociometric data. Research such as \cite{olguin2009capturing} chose the period of a day as a meaningful unit, and proposed that differences in behavior on different days will reflect changes in team performance. The choice of five-minute slices by Pentland was useful for demonstrating the value of shorter periods of data, but the choice was seemingly arbitrary. An alternative approach is to use the data itself to search for meaningful divisions of time. In exploratory research by Zhang et al. \cite{zhang2018team}, as a complement to their predictive work, the authors applied topic modeling techniques and found that recordings of social behavior could be reduced to lower-dimensional representations reflecting the routines of participants in the simulated space missions. This suggests that sociometric data can exhibit its own temporal structure, something that we attempt to exploit. We apply dynamic complexity, a method from the analysis of complex systems in order to segment sociometric data into periods representing different phases of interaction. To the best of our knowledge, we are the first to apply the method of dynamic complexity to data from sociometric badges and to visualize structural changes in networks based on this approach
~\cf~\cite{guimera2005team}. 

\subsection{Analysis of Social Interaction Networks}
Prior work has shown that the strength of connections in networks can be an important predictor of collaborative team performance \cite{de2014strength} and that more dense network connections are associated with teams that are better able to accomplish their goals and persist as a cohesive team \cite{balkundi2006ties}. Such networks can be estimated with respect to \eg using online social interactions as well as physical (offline) interactions, \cf~\cite{Atzmueller:14:CoRR}.
%%Online interactions are typically derived from data on specific online systems, \cf~\cite{MABHS:11,MASH:13} for a discussion using Twitter and Flickr.
For the latter, sociometers are applied from which one can then estimate and model offline social interactions between people~\cite{Atzmueller:14:CoRR}. Indeed, this is the direction we focus in this paper.

As one of the first approaches to modeling offline social networks, Eagle and Pentland~\cite{eagle2009inferring} utilized proximity information estimated by Bluetooth devices as a proxy for human interaction. While approximating face-to-face communication, proximity may not always directly reflect face-to-face interaction~\cite{BCCPBV:08}. Also, proximity does not capture vocal and/or (non-)verbal aspects of communication. For this reason, the Sociometric Badge, and its more modern successors (\ie the OpenBeacon tag and the Rhythm Badge), were adopted for understanding offline social networks as they record more information about the interaction. In this paper, we apply the Rhythm badges~\cite{lederman2018rhythm}, as they provide a richer set of information regarding social analytics of teams compared to the Openbeacon sensors. 

In our work, we do not target traditional network modeling of these data, but instead focus on the computational sensemaking using visualization-based approaches. Notably, Pentland proposed that sociometers could be used to provide a scientific approach to teamwork \cite{pentland2012new} by his observation that teams performing well on a variety of objective measures and also subjective measures can be distinguished from low performing teams by observing the patterns of communication between team members. Our work focuses on two of the measures proposed by Pentland that can be calculated from vocal activity recordings using the Rhythm badges. These measures represent the ‘energy’ and ‘engagement’ of team members and can readily be visualized in network form. However, prior work is limited so far in its capacity to detect critical changes to team interactions over time.

\subsection{Critical Instabilities and Phase Transitions}
Critical instabilities, sometimes referred to as critical fluctuations or phase transitions, are pervasive across physical, biological, and ecological systems \cite{kelso1986nonequilibrium,scheffer2012anticipating}. The core dynamics behind critical instabilities is change over time that marks the transition of system states such as the shift of water from liquid to vapor, an ecosystem shift from one stable state to a another, a shift in quadrupedal locomotion from walking to trotting, and a change in group interaction processes. Phase transitions, more generally, are theorized to mark structural changes in the organization of system components that can happen gradually or suddenly in a non-linear fashion \cite{schiepek2010identification}. Here we use critical instabilities as indicators of points that a phase transition has likely occurred. Such transition points are key to identify and examine given that systems at, or approaching, transitions are in periods of shifting instability and are thus, highly influenceable \cite{thelen1991hidden}. 
So far, however, examining phase transitions in team research has been limited. Wiltshire and colleagues \cite{wiltshire2018problem} used an entropy-based method to detect phase transitions in the collaborative problem solving communications of dyadic teams performing a complex, computer-mediated task. They were not only able to empirically identify phase transition points that separated distinct communication process distributions, but they also observed that lower entropy values at transition points were associated with better team performance. Amazeen, Likens et al. \cite{likens2014neural,amazeen2018physics} also used an entropy-based method to look at changes in the neurodynamic \cite{likens2014neural}, communicative \cite{gorman2016cross}, and physiological \cite{dias2019physiological} organization of teams. They found that they were able to recover, without any prior information, changes in the interaction based on the task context (\eg moving from training to performance period) and key events (\eg a fire in a simulated operating room), and that more experienced teams were more efficient at undergoing the critical transitions in the communicative interactions \cite{gorman2019measuring}.

While there are many measures that could be utilized to attempt to identify critical instabilities and phase transitions \cite{dakos2012methods}, dynamic complexity \cite{schiepek2010identification} is one measure that holds particular promise for team communication data and examining multi-modal interactions, given (1) its ability to provide valid results on short and coarse grained time series (minimum of seven data points per window and time series length of 20 observations), and (2) as it can be applied to any interval-scaled and regularly sampled data with no assumptions and then, visualized as a heatmap to highlight points of critical instability across multiple measures \cite{schiepek2014self}. Prior work utilizing measures of dynamic complexity, while not in studies of teamwork, have shown that during psychotherapy, dynamic complexity was useful for identifying critical fluctuations in patient self-reported symptoms that preceded key clinical changes \cite{olthof2019destabilization}. Given the ability of dynamic complexity to identify critical transitions, we expect that when applied to sensor data from teams we would be able to detect key transitions in tasks and interactions, that may also correspond to transitions in teams' network structure \cite{guimera2005team}.

\subsection{Overview of Current Study}
In this study, we present a case study of a single team performing a collaborative task as a proof of concept for the scientific novelty and utility of the unique combination of theory, sensors, and visualizations. By combining theory on phase transitions with Rhythm Badge data, we aim to improve research on team dynamics by showcasing the combined use of dynamic complexity heatmaps and network visualizations for team communication data. 
We predicted that dynamic complexity would be able to recover known task transitions as well as unknown event/interaction transitions. And, further, that we would observe changes in network structure in the phases detected between transition points. 

\section{Method}
\subsection{Participants}

For this case study, a group of seven people voluntarily participated. They comprise the management team of a large tech company based in the Netherlands, and thus, were operating in their natural team constellation. All participants had significant working experience: 10-14 years (29\%),  15-24 years (57\%), and $>$25 years (14\%). Three participants were working together in this specific team between 8-14 years, one between 4-7 years, two between 1-3 years and there was one participant that joined in the last year. Two participants were female (29\%) and 5 male (71\%). Their age groups ranged from 25-34 (29\%), 35-44 (29\%) and 45-55 (42\%). 6 participants had either a Bachelor (29\%), Master (57\%) or equivalent as their highest education level, and one participant (14\%) indicated ‘other’.

\subsection{Materials}
Participants were seated at a table all facing each other, like in a typical meeting setting. Each participant was required to wear two sensors – the Rhythm Badge sensor to measure vocal amplitude and proximity, as well as an OpenBeacon sensor to also measure proximity, only at a higher sample rate than the Rhythm sensors (OpenBeacon data was not used in this study).
Audio data from the Rhythm Badge is recorded by a microphone at a rate of 700Hz and split into 50ms sections and then averaged. Voltage is also recorded, which can help to identify badges that run low on battery, but this has little value for measuring social behavior. In this paper, we focus on the vocal amplitude measure for dynamic complexity analysis.

A laptop in combination with a Raspberry Pi was used to run the data collection protocol for the Rhythm sensors. The experiment was also captured using video to be able to verify output resulting from the data analyses. All participants received an answer sheet they needed to complete during the experiment. Before and after the experiment the participants were asked to fill out a total of three questionnaires (listed below), the output of which will be used in future research as they were beyond the scope of the current study. 

\subsection{Procedure}
Prior to the experiment, the participants were asked to fill out two questionnaires. One containing demographic as well as questions related to their current team behavior and environment covering themes like team learning behavior, psychological safety and alignment and the other was a personality profiler \cite{minnee_verberg_2012} to establish the preferred thinking and action style of each participant. 

For this study, the Lost at Sea paper-based simulation was used \cite{nemiroff2001lost}. This task was selected because our goal was to uncover team dynamics in a meeting-like setting where group decision-making and seeking consensus is central. The team was told their boat had caught fire in the middle of the Atlantic Ocean. They were able to make it to a life raft with 15 items and a box of matches. Participants were first asked to individually rank these items in order of importance to assure the highest chance of survival. Next, participants were asked to discuss their rankings and reach a group consensus of the team's collected item ranking. The optimal order and thus correct answer was based on the expert ranking given by the US Coastguard. The difference in order for each item is summed to calculate the total score. The lower the score, the better the team performed. 
There was no time constraint for the task, but a competitive and motivational element was introduced by informing participants that they were taking part in a larger study and that the overall scores of each team will be compared such that the team with the highest score can win a ‘prize’. Overall rankings were determined by taking into account not only the difference in score between the expert ranking and the team ranking, but also how long it took the team to reach consensus. 

After the experiment, but before sharing of the expert ranking, participants completed a short questionnaire where they were asked to indicate their emotional state during the experiment \cite{warr1990measurement} as well as how they would rate their team’s output. They were also asked to assess if they behaved like they would in a similar real-life situation, which  scored an average of 4.9 on a 7-point Likert scale. 

\subsection{Data Processing} 
Several data processing steps were involved.\footnote{The routines used to process the data and create the visualizations were written in Python 3.7.3 \cite{van2007python}, with the numpy \cite{walt2011numpy} version 1.16.2 and pandas \cite{mckinney2010data} version 1.0.3 packages used for data manipulation. The seaborn package \cite{michael2018seaborn} version 0.9.0 was used to create the dynamic complexity heatmaps. }
An important initial step involved in processing the data from the Rhythm Badges is voice activity detection (VAD). We use a correlation-based algorithm provided by MIT in a package of analysis routines for Rhythm \footnote{\url{https://github.com/HumanDynamics/openbadge-analysis}}. The algorithm operates on the volume signals measured from the Rhythm badges. Using the correlation between signals, it aims to distinguish moments when the individual was speaking from moments when they were silent but their badge recorded other speakers (a phenomenon known as 'cross-talk') or background noise. The correlation threshold was set to .40 as this has been shown to be effective in testing VAD. The output is a list of individuals who were speaking during each second of the audio recording. 
Full details of the algorithm are provided in Appendix B of \cite{lederman2019founders}. This method for voice activity detection was chosen because it is capable of identifying multiple simultaneous speakers and because it can be readily applied to data in the form used by Rhythm badges. The time series extracted by VAD was used as an input for the network visualizations. 

During the experiment, one of the badges (worn by participant C) stopped recording. The data that was collected by this badge was included in the data processing, however the fact that there were periods of no recording should be kept in mind when interpreting the results presented below. Normally it would be expected that interactions between participant C and other participants would be reflected in the network visualisations, however due to the missing data these interactions are not shown in all of the visualisations in the Results section. 

\subsection{Dynamic Complexity Analysis}

In order to identify critical instabilities, points at which the dynamics of the interaction between team members change, we use a dynamic complexity based approach. Dynamic complexity is a measure of how complex the behavior of a system is, and its use for detecting critical instabilities is guided by the insight that phase transitions between one pattern of behavior and another, within a complex system, are heralded by brief periods of highly complex behavior \cite{schiepek2010identification}. The dynamic complexity measure applies to time series and is calculated within a moving window, quantifying the complexity exhibited by the system over time. Once dynamic complexity values have been computed for a time series, it is possible to search for moments of unusually high dynamic complexity, which indicate a critical instability and an associated phase transition. 

Dynamic complexity is calculated by combining two components called the 'Fluctuation' and the 'Distribution', both of which are also calculated over a window of values. Both values range from 0 to 1. These quantities are originally presented in \cite{schiepek2010identification}.

\textit{Fluctuation} ($F$) is based on changes in value between ‘points of return’ where the gradient stops being positive, zero, or negative (i.e., looking in between these points, we see periods when the value is consecutively increasing, staying the same, or decreasing). For each period between points of return, the difference between values at the beginning and end of the period is divided by the number of time points included in the period, and the outcomes are summed across the entire window. The result is then divided by a ‘maximum possible fluctuation’ for the window size, which is what would happen if there were only oscillation between the maximum and minimum values. $F$ is high when there are frequent oscillations between high and low values, and is lower when oscillations are less frequent or have smaller amplitude. 

The Fluctuation $F$ for a window with size $m$ is formalized as follows:
With a window $w = (x_1, \ldots, x_n)$ containing a sequence of $n$ values $x_i \in X$ for a domain $X$ and the index set $I = \{1, ..., n\}$ and a subsequence of "points of return" including the start and end point of the window $p = (x_j), j \in P, P \subseteq I$, where a "point of return" is defined as above and a function $\phi: X \times P \rightarrow I$ which maps an element contained in the subsequence $p$ to its original position (index) in the window $w$ 
(\eg if the $kth$ element $p_k$ contained in P originally corresponds to the $jth$ element of $w$, \ie $x_j$ then $\phi(p_k) = j$),
we compute the sum of the absolute differences between subsequent elements in $P$, normalized by the differences of their respective indices in $I$, and normalized according to window size $(m - 1)$ and maximal difference in values $d = (x_{max} - x_{min})$, with $x_{max}$ and $x_{min}$ being the maximum and minimum values in the domain $X$. Now, for the Fluctuation $F$ we obtain:

\begin{equation}
F = \left(\frac{1}{d(m - 1)}\right)
    \sum_{k=1}^{|P|-1} \frac{|p_{k+1} - p_{k}|}
                            {\phi(p_{k+1}) - \phi(p_k)}
\end{equation}

\textit{Distribution} ($D$) compares how regular the differences are between sorted values in a window. First, the values in the window are sorted in ascending order. Then, the sorted observed values are compared to artificial series of values that were all evenly-spaced across the possible measurement scale. The disparity in the differences (e.g., in the differences between the first and third values within each sorted list) is the basis for $D$. Positive disparities are summed. This addition continues to be calculated across the whole window for different sub-window sizes. So, first, the differences at an offset of 1 (e.g., first and second sorted values) will be considered, followed by differences at an offset of 2 (e.g., first and third sorted values). Before being added into the summation, each difference is normalised by dividing it by the evenly-spaced expectation of what the difference should be. $D$ is highest when there is an equal distribution of values, and it decreases when there is a preference for certain values or certain ranges of values. 

The Distribution $D$ is then given as follows:

\begin{equation}
D = 1 - \sum_{c=1}^{m-1}\sum_{d=c+1}^{m}\sum_{a=c}^{d-1}\sum_{b=a+1}^{d} \frac{\Delta_{ba}\Theta(\Delta_{ba})}{\delta_{Y_{ba}}} \,,
\end{equation}

where $m$ is the size of the window; $\Delta_{ba}$ is the difference between the values at $a$ and $b$ within the sorted values; $\Theta$ is the Heaviside step function which gives $1$ when the input is positive and $0$ otherwise; and, $\delta_{Y_{ba}}$ is the difference between the values at $a$ and $b$ within the artificial values. 

\textit{Dynamic complexity} within a given window is simply the product of the F and D measures. We calculated dynamic complexity upon each of the time series derived from the volume recorded by the microphones of the Rhythm Badges. The Rhythm Badges themselves produce a time series in which each data point is the average volume recorded over a 50ms period. We resampled this series to a rate of 0.2 Hz and took the average within each 5s period. This was done for two reasons: first, it reduces the size of the time series and so lessens the computational demand of the process, and second, it lowers the impact of fluctuations in volume that occur naturally in speech (for example due to pauses for breath) that are incidental and not indicative of changes in team dynamics. To calculate the dynamic complexity, we used a sliding window size of 12 with a increment of 1, which corresponds to a one-minute period. 

We then calculated the average dynamic complexity across the time series for all team members. This resulted in a single series of dynamic complexity values from which critical instabilities can be identified by moments of high complexity. Since local maxima of complexity are indicative of critical instabilities \cite{schiepek2010identification,olthof2019destabilization,thelen1991hidden}, we emphasize local maxima over the global maximum. We use a moving window within which dynamic complexity values larger than two standard deviations from the mean for that given window will be identified as points of critical instability. For this purpose, we use a window size of 60, which corresponds to the complexity scores encountered over a five-minute period. When then use the communication time series falling between these phase transition points as input for the network visualizations.

\begin{figure*}[htb]
    \centering
    \includegraphics[width=0.30\textwidth]{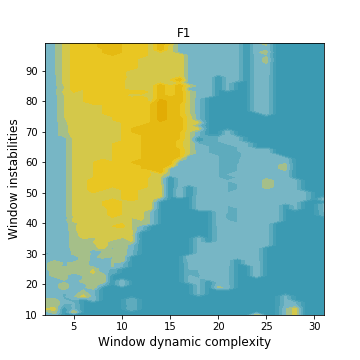} \includegraphics[width=0.30\textwidth]{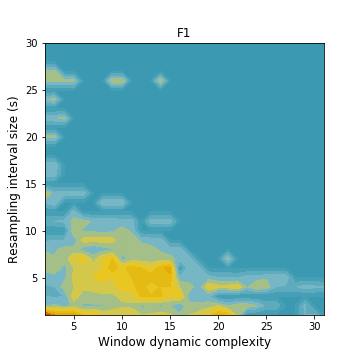} \includegraphics[width=0.375\textwidth]{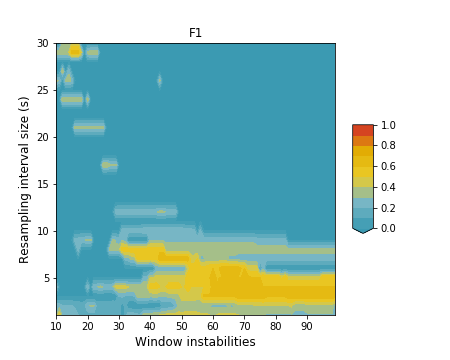}
    \caption{Pairwise investigations of parameter settings}
    \label{fig:pairwise_explorations}
\end{figure*}

\subsection{Network Visualization}

Following the identification of critical instabilities, we use two measures capturing different communication patterns that not only reflect the quality of the social interaction, but that also help us to better understand the team dynamics in a visual manner. This approach is motivated by Pentland and colleagues' work \cite{pentland2012new} in which they posited that high-performing teams could be identified by certain characteristics of their communication patterns. Unfortunately, precise details of how the measures were calculated are not available (to the best of our knowledge) and so our approach, described below, is inspired by, rather than an exact match to the measures advanced by Pentland and colleagues.

The first measure is the \textit{individual utterance rate}, which is analogous to Pentland's 'energy' measure \cite{pentland2012new}. It reflects how much each individual is contributing to the overall communication for a given time window. Any continuous period of speaking, as identified by VAD, is considered an utterance. The rate is the number of utterances started divided by the duration of the time period (in seconds). In our visualizations, each team member is represented by a node and their energy (relative to other team members) is depicted through the size of the node. 

The second measure used is the \textit{pairwise rate of responses between two individuals}, which is analogous to Pentland's 'engagement' \cite{pentland2012new}. A response is considered to occur when the first individual begins an utterance within five seconds of the second individual speaking. The rate is the number of times this occurs in the time period divided by the length of the time period. This measure gives an indication of which team members are interacting with which other team members, and which people are involved in the same discussions. In the network visualizations, the rate of responses is portrayed as the thickness of the edge connecting two nodes. 

The measures described above can be combined to summarize a period of interaction in a single image. In prior work balanced networks, where all nodes and edges have similar sizes, have been argued to be indicative of high-performing teams \cite{pentland2012new}. By contrast, network imbalances could be used to identify problems such as team members who are overly dominant or team members who are seemingly excluded. Examples of these network visualizations are presented and discussed in more detail in the Results section.

\subsection{Exploration of Parameter Settings Viability}

Recall that, we resampled the raw volume data with a resampling interval of five seconds, creating a 0.2Hz time series. We used a window size of 12 to calculate dynamic complexity and a window of 60 to derive critical instabilities, corresponding to a one-minute and a five-minute period respectively. While we chose the aforementioned default parameters based on prior work and careful considerations of the data, we systematically explored these parameter selections to 1) empirically evaluate the robustness of these parameters for our case study data and 2) inform future efforts on sensible parameter settings.

To systematically investigate the parameters, we performed pairwise comparisons, varying the value of two parameters while keeping the remaining parameter fixed at its default value. This makes it possible to examine the interactions between pairs of parameters as well as their individual effect. The performance of parameter settings was evaluated by treating the derived instabilities as predictions and known task transitions as the correct labels, then computing F1 scores.

Figure \ref{fig:pairwise_explorations} illustrates the F1 scores of three pairs of parameters, using contour maps with 10 levels. Warmer colours indicate a higher F1 score. From the plots, the following parameter regions are sensible: 

\begin{itemize}
    \item Window to calculate dynamic complexity: between 10 and 15
    \item Window to identify critical instabilities: between 40 and 100
    \item Resampling interval size: between 3 and 6 seconds
\end{itemize}

In addition, we ran a three-dimensional grid search to examine the interactions between all three parameters to determine if there was a better combination of settings. It suggested that slight improvements in the F1 score could be made by increasing the resampling interval to either 9 or 14 seconds while reducing the other two parameters. However, one drawback is that these parameter combinations require window sizes of only 7 or 8 for dynamic complexity, which is at the minimum considered reliable. The three-dimensional grid search also found that the following combination gave the highest F1 score: window size for dynamic complexity equal to 20, with the resampling interval of 4 seconds and window sizes of 80-85 for finding critical instabilities. The issue with this combination, and combinations with resampling intervals of 9 or 14, is that they appear in small regions of successful settings, implying that they are more unstable in the face of variation, or that they might only work in this instance due to the peculiarities of our case study dataset.

Taken together, these results indicate that the initial choice of parameters was reasonable, that using other parameters would not provide substantial benefits that outweigh the disadvantages of those parameters, and that the settings produced stable results (i.e., small changes to parameters do not drastically reduce the quality or quantity of the derived instabilities). 

\subsection{Reliability of Voice Activity Detection Method}

The network visualisations used in our method are based upon turn-taking behaviours, which themselves rely on the outputs of a voice activity detection (VAD) algorithm that aims to identify who spoke and when. Since the VAD processing step can potentially impact the network visualisations, we investigated the reliability of the VAD outputs by comparing them to manual annotations of the video recording.

Manual annotations were created for a subset of the video recording, consisting of 460 seconds from Team Task segment of the experiment. Annotations were recorded using the ELAN software system\footnote{From the Max Planck Institute for Psycholinguistics, The Language Archive, Nijmegen, The Netherlands: \url{https://archive.mpi.nl/tla/elan}} \cite{elan}. Inter-rater reliability scores, in the form of Cohen's kappa and Gwet's AC1\footnote{Past work \cite{wongpakaran2013comparison,zec2017suppl} suggests that Gwet's AC1 is less biased by a 'prevalence effect', where one outcome of a variable is noticeably more common than the others. Since 'non-speaking' was more common than 'speaking' in our data, a prevalence effect may impact these scores.}, were then computed from the manual annotations and the outputs of the VAD algorithm for each team member. Across the team, Cohen's kappa scores varied from .32 to .65, with a median value of .47, while Gwet's AC1 varied from .29 to .82, with a median of .62. The lowest inter-rater reliability according to both metrics was for an individual to whom the manual annotator attributed more speaking time. To establish the reliability of the manual annotations, we also compared the manual annotator to two additional manual annotators, on samples of roughly 90 seconds from the Team Task. Inter-rater reliability for the first additional annotator ranged from .43 to .87 with a median of .73 for Cohen's kappa, and from .66 to .95 with a median of .74 for AC1. With the second additional annotator, Cohen's kappa ranged from .72 to .84 with a median of .78, and AC1 varied from .75 to .90 with a median of .85. Overall, these results suggest that manual annotations generally have good levels of inter-reliability, and that the VAD algorithm has moderate reliability compared to human annotation, although there is some variability across team members.

\begin{figure*}[htb]
    \centering
    \includegraphics[width=1.0\textwidth]{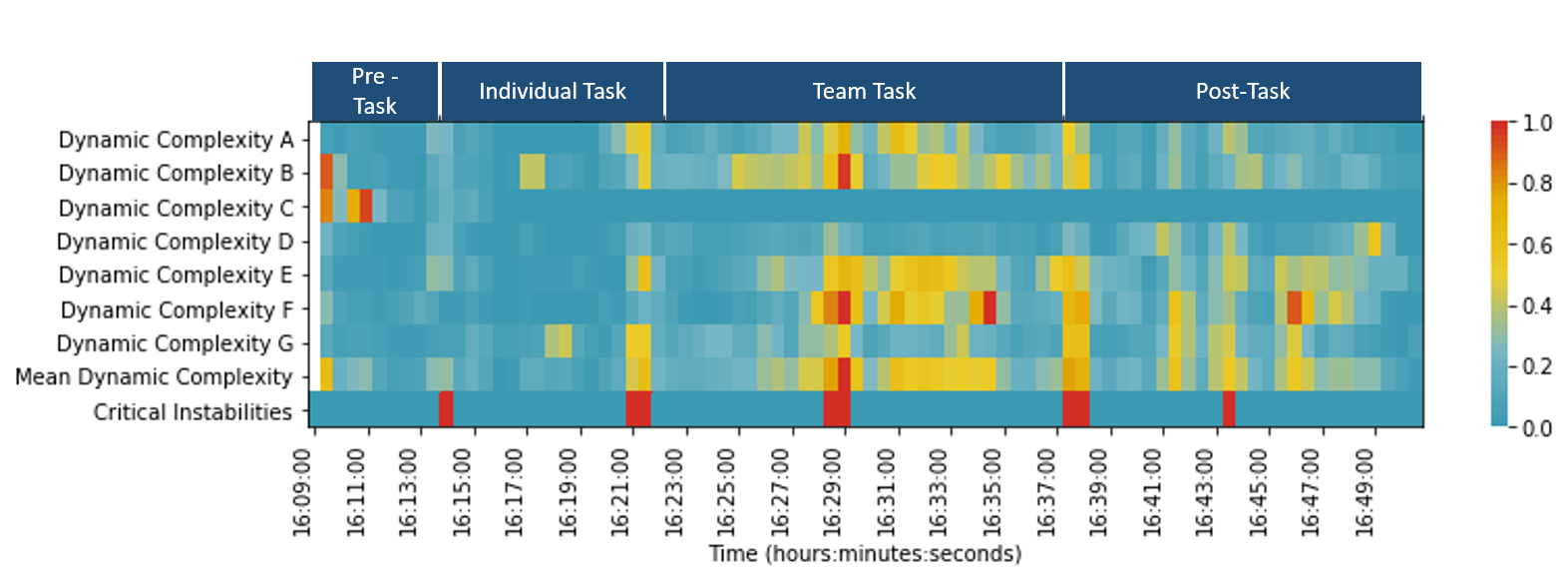}
    \caption{Dynamic complexity heatmap}
    \label{fig:heatmap}
\end{figure*}

\section{Results}
In order to meet the aims of our study, the results are structured as follows. First, we start with the Dynamic Complexity heatmap to highlight periods of critical instability. We then compare these to known task transitions and report on investigations of the videos regarding what occurred around those time points. Next, we highlight the network visualization for each of the phases identified using the dynamic complexity heatmap.

In Figure \ref{fig:heatmap}, the dynamic complexity for each of the seven team members' voice level occupies the first seven rows. Below this is the average dynamic complexity across all team members, and then the time-localized instances of critical instability identified using our method described previously. Along the x-axis, times are listed over the duration of the experiment. The values of dynamic complexity are cooler when low (blue) and warmer when high (red). 

By identifying time-localized critical instabilities at both the individual level (team member rows) and at the team level (the Critical Instabilities row), we gain insight into the dynamic structure of the team's interaction. Most notably is the fact that, as predicted, dynamic complexity was sensitive to known task transitions. The known segments of the study are shown at the top of the figure (e.g., Pre-Task, Team Task, etc.). If we compare the time points for known task transitions at the top of Figure \ref{fig:heatmap} to those periods of instability highlighted at the bottom of the figure, we see we are able to recover all these transitions. For example, we can clearly see the transition from completing the collective ranking (Team Task) immediately following 16:37:00. Furthermore, as expected we also identified points of instability within task transitions (those that were not known prior to the analysis and correspond to changes in the interaction). We observe this during the both the Team Task (16:28:30) and during the Post-Task section (16:43:30). 

At this point, it is apparent that the combined use of Sociometric sensors that capture vocal volume of team members and the visualization of dynamic complexity with critical instabilities identified, provides an empirically derived and objective overview regarding the team dynamics. This is particularly insightful given the ability of the measure to capture the identification of task- and interaction- transitions. To make better sense of the critical instabilities and interaction transitions, two members of our research team evaluated the video recording as well as notes of the interaction session. Table \ref{table:qualitative} provides a detailed overview of the times in the task when team critical instabilities were identified and qualitative assessments of what occurred at those points as well as the period in between them. With respect to the interaction transition during the Team Task, the critical point corresponded to a team member who previously had not contributed much proposing an unpopular perspective that changed the dynamic. Many sub-conversations ensued. Regarding the Post-Task critical point, the facilitator began to provide the expert rankings (solution to Lost at Sea task) and many participants expressed excitement. So this represented a critical event in the interaction. 

\begin{table*}[h]
\begin{center}
\begin{tabular}{ |p{0.12\textwidth}|p{0.38\textwidth}|p{0.44\textwidth}|  } 
    \hline
    \textit{Time of Critical Instability} & \textit{Qualitative Description of Critical Instability} & \textit{Qualitative Description of Period Until Next Critical Instability} \\
    \hline
    16:09:00 & Facilitator gave an overview of the experimental session. & \\
    \hline
    16:14:00 & Individual ranking part of the task begins.  & Most participants are quiet and focusing on completing the task. Some side discussions develop between participants after they complete their task. \\
    \hline
    16:21:00 & Transition from individual ranking to team ranking with a brief period of discussion and an explanation from the facilitator.  & While there is some discussion between the entire group during this period, two participants primarily take the lead during this beginning part of the team ranking discussion. \\
    \hline
    16:28:30 & Interaction dynamic changes as a previously quiet team member introduces a novel viewpoint and multiple sub-conversations ensue.  & Participants who were less active in the previous phase are now more active \\
    \hline
    16:37:30 & The team consensus has been reached and the timer is stopped. The group gets quiet and the facilitator asks participants to fill out the questionnaire.  & Some members are focusing on completing the questionnaire. Other members ask questions and engage in dialogue. \\
    \hline
    16:43:30 & The facilitator begins to provide the expert ranking of the items, and the team expresses excitement when the first item is announced.  & As each of the expert rankings of the items is provided by the facilitator, the group discusses the item and makes jokes. Given the correct rankings, individuals are silent as they assess their individual scores. \\
    \hline
\end{tabular}
\end{center}
\caption{Qualitative description of critical instabilities and periods between critical instabilities}
\label{table:qualitative}
\end{table*}

\begin{figure*}[t]
    \centering
    \includegraphics[width=0.95\textwidth]{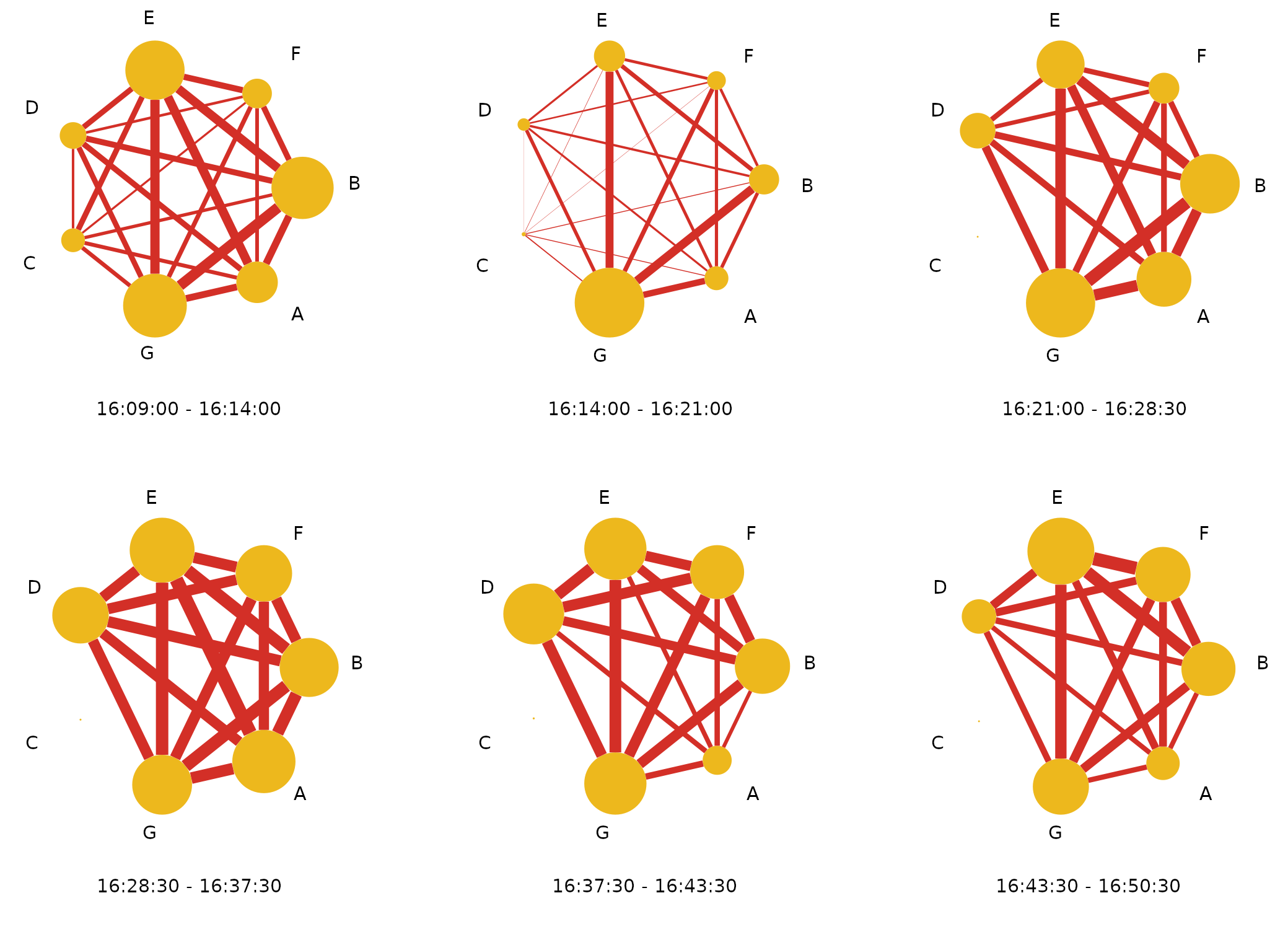}
    \caption{Network visualizations during periods between critical instabilities}
    \label{fig:network_all}
\end{figure*}

Next, for each of the identified phases of the experiment, we provide visualizations of the network analysis described previously in Figure \ref{fig:network_all}. Recall that localized peaks in dynamic complexity should signify changes to the structural organization of the team. Thus, we expected to observe concomitant changes in the network visualizations and, for the most part, the visualizations confirm this. Across the phases of the task, we see changes in the amount of energy of each speaker as well as the engagement. In some cases, the changes are subtle such as the difference between the phases of the Post-Task (16:37:30-16:43:30 \& 16:43:30-16:50:30), where participant D exhibits a decrease in energy and engagement with many of the other participants. In other cases, the changes in the network visualization are more drastic such as the difference between the Individual Task (16:14:00-16:21:00) and the Team Task (16:21:00-16:28:30) where engagement and energy from nearly all members is increased. Thus, from a visual analytics perspective, there do appear to be structural changes in the interaction dynamics of the team that can be readily visualized in this format. 

\section{Discussion}
Many foundational studies of team work have painstakingly worked to identify critical team work events that lead to incidents or breakdowns in team coordination often by manually analyzing videos, transcriptions, and conducting interviews, \eg \cite{bearman2010breakdown,klein1989critical}. In our work, we provided a novel approach using Sociometric badges that quickly captures team communication information, and that provides a visual analysis of the dynamics of the team by showcasing the changes in dynamic complexity for individual members and the team as a whole. This allows for an at-a-glance interpretation of the changing team dynamics and identification of critical transition points in the team. We took this one step further by visualizing the changes in team member energy and engagement across the phases of this case study (identified by critical instabilities), using network visualizations. Combined, these measures, analyses, and their visualization have potential to provide important insights on team dynamics across many domains including business, healthcare, science, politics and governance, as well as aviation and spaceflight.

While we present a novel case study here, this work is situated in the broader, albeit relatively new, area of work using Sociometric badges to understand team work dynamics (e.g., \cite{zhang2018team,kozlowski2018unpacking}). That being said, given that this is a case study, there are obvious limitations to the current work, which we expect can be addressed when when scaled up to a full experiment. This case study has demonstrated the potential of an approach based on dynamic complexity and network visualisations, but future work is needed to establish the robustness of such an approach across multiple contexts and experimental sessions. Moreover, examining the relationships between patterns of dynamic complexity, points of critical instability, and networks with team performance, personality, as well as well-being measures is a crucial next step that a full experiment could address. 

There are many potential directions to pursue in this regard. Importantly, future efforts should work toward understanding the underlying factors that contribute to the \textit{optimal} team dynamics that lead to excellent performance and teams with high well-being. A challenge with phase transition detection methods as well as network visualizations is that we can observe a critical change in the dynamics, but their precise meaning is elusive. By combining our approach with other measures such as questionnaires capturing psychological safety, accountability, alignment, and focus as well as the cognitive diversity within teams using personality measures, \eg \cite{minnee_verberg_2012}, we expect that researchers and practitioners can begin to better illuminate the meaning of changes in team interactions.

The inclusion of other measures of teamwork could also provide a sound basis for investigating the impact of pre-processing on the analysis of sociometric data. For example, the results that arise from different choices of VAD algorithm and VAD parameter settings could be validated against other measurements, providing an insight into how to tune the end-to-end visualization procedure. 

Another direction, when considering that Pentland \cite{pentland2012new} argued that balanced networks might be important for different types of performance (e.g., creative teams and feeling heard), is that we expect that this is highly contingent on task context. One approach in the future might be to utilize a form of pattern classification of the networks on the engagement and energy levels of team members such as those utilized by Stevens and colleagues \cite{stevens2019teaching} to identify robust network structures that have a clearer meaning such as whether these patterns relate to cohesion, effective collaboration, or team well-being. Extending this even further to include additional modalities of the interaction could also be a possible future step~\cite{paxton2013multimodal}.

In future work, we aim to include other modalities into the investigation, \eg by considering proximity information, \eg~\cite{BCCPBV:08,Atzmueller:14:CoRR} and data from online social interaction networks~\cf \cite{MASH:13}, as potential complementary information providing further insight on multi-modal and multi-scale team interaction. Methods for modeling and analyzing feature-rich multiplex networks
~\cite{IAGKLS:19} are promising directions.

\subsection{Practical Implications}
That communication and team interaction have an impact on task performance \cite{marlow2018does} and well-being \cite{rosenfeld1997developing} is not debated in academia or in practice. However, both domains often struggle with measuring and visualizing the quality of team dynamics and the transition points that contribute (either positively or negatively) to these dynamics. Using the social analytics approach employed here, practitioners can more readily derive insights. The aim is that these methods could begin to make it more clear and transparent what teams ought to change in their behaviors, attitudes, and cognition to reach more optimal team dynamics, that, in turn, result in a high-performing and happy team. 

With the methods demonstrated in this case study, one can now objectively analyze the dynamics of a team, pinpoint the critical moments, and give instant and intuitive feedback to teams through visualizations. This is particularly relevant for many time sensitive domains and those focusing on trying to implement real-time feedback for team performance improvement \cite{dias2019physiological,gorman2012measuring,wiltshire2020challenges}. This visual feedback is a valuable application to improve the team dynamics because it allows team members to objectively see their role within the team, such as how much energy or engagement they contributed, and how this effects the overall team’s dynamics. Such visualization-based feedback may result in increasing rates of acceptance and accountability for individual behavior that can be a crucial element to eliciting behavioral change.

By measuring team dynamics over a longer period and utilizing these techniques, the effect of certain interventions (e.g., new working methods, new team composition, and behavioral agreements) on the team dynamics may become more transparent. We anticipate that it could also facilitate an iterative visual feedback and learning cycle by teams. Not only this, but by having the possibility of identifying critical transition points that correspond to changes in team dynamics, teams can also begin to learn how to either mitigate negative or stimulate positive events. 

\subsection{Conclusions}

In conclusion, we have contributed what we see as a foundational case study for a social analytics approach to team dynamics by combining sociometric sensing, dynamic complexity heatmaps, and energy and engagement network visualizations. We think this method holds much promise and potential for future scientific and practical work aimed at improving the performance and well-being of teams in many of today's complex sociotechnical work domains.

\section*{Acknowledgements}

We kindly thank Oren Lederman for helpful discussions and for granting us access to the use the Rhythm badges. Additionally, we are grateful for the contributions of Bo (Ngoc) Doan's assistance in testing and implementing the Rhythm badges and contributions to the study setup. This work was partially funded by the Dutch Research Council (NWO) as part of the NWO-KIEM Creative Industries \& Digital Humanities program, project number KI.18.047. The research leading to this work has also received funding by the German Research Foundation (DFG) project ``MODUS'' under grant AT 88/4-1.

%%%\bibliographystyle{spmpsci}
%%%\bibliography{references}

\end{document}